\documentclass[prb,aps,showpacs,twocolumn,unsortedaddress,superscriptaddress]{revtex4-1}
\usepackage{graphics, bm, psfrag, amsmath, amssymb, epsfig, grffile, float}
\usepackage{color}
\usepackage{booktabs}
\usepackage{tabularx}

\begin{document}

\title{Excitonic Gap Formation in Pumped Dirac Materials}
\author{Christopher Triola}%~\footnote[1]{These authors contributed equally to this work}}
\affiliation{Nordita, Roslagstullsbacken 23, SE-106 91 Stockholm, Sweden}
\affiliation{Center for Quantum Materials (CQM), KTH and Nordita, Stockholm, Sweden}
\author{Anna Pertsova}%~\footnotemark[1]}
\affiliation{Nordita, Roslagstullsbacken 23, SE-106 91 Stockholm, Sweden}
\affiliation{Center for Quantum Materials (CQM), KTH and Nordita, Stockholm, Sweden}
\author{Robert S. Markiewicz}
\affiliation{ Physics Department, Northeastern University, Boston MA 02115, USA}
\author{Alexander V. Balatsky}
\affiliation{Institute for Materials Science, Los Alamos National Laboratory, Los Alamos New Mexico 87545, USA}
\affiliation{Nordita, Roslagstullsbacken 23, SE-106 91 Stockholm, Sweden}
\affiliation{Center for Quantum Materials (CQM), KTH and Nordita, Stockholm, Sweden}
\affiliation{ETH Institute for Theoretical Studies, ETH Zurich, 8092 Zurich, Switzerland}

\date{today}

\begin{abstract}
Recent pump-probe experiments demonstrate the possibility that Dirac materials may be driven into transient excited states describable by two chemical potentials, one for the electrons and one for the holes. Given the Dirac nature of the spectrum, such an inverted population allows the optical tunability of the density of states of the electrons and holes, effectively offering control of the strength of the Coulomb interaction. Here we discuss the feasibility of realizing transient excitonic instabilities in optically-pumped Dirac materials. We demonstrate, theoretically, the reduction of the critical coupling leading to the formation of a transient condensate of electron-hole pairs and identify signatures of this state. Furthermore, we provide guidelines for experiments by both identifying the regimes in which such exotic many-body states are more likely to be observed and estimating the magnitude of the excitonic gap for a few important examples of existing Dirac materials. We find a set of material  parameters for which our theory predicts large gaps and high critical temperatures and which could be realized in future Dirac materials. We also comment on transient excitonic instabilities in three-dimensional Dirac and Weyl semimetals. This study provides the first example of a transient collective instability in driven Dirac materials.
\end{abstract}


\maketitle
\section{Introduction}
Dirac materials (DMs) represent a growing class of systems including superfluid $^3$He, high-temperature $d$-wave superconductors, graphene, and the surface states of three-dimensional topological insulators (3DTIs)~\cite{wehling2014dirac, dahal2010charge, fu2007topological_1, fu2007topological_2, zhang2009topological, zhang2014proximity}. The defining feature of a DM is the existence of Dirac nodes in the low-energy excitation spectrum leading to an energy-dependent DOS which vanishes exactly at the Dirac point, e.g. $\mathcal{N}(E)\propto E$ for two-dimensional (2D) DMs such as graphene and 3DTI surface states. The disappearance of the DOS at the nodal point leads to a critical coupling for many-body instabilities which can gap the spectrum~\cite{kotov2012electron}. Several previous studies have investigated the phase diagram of DMs with respect to the material-specific fine structure constant, $\alpha$, and suggested a critical value of this constant, $\alpha_\mathrm{c}\approx 1$, above which the material is expected to be an excitonic insulator~\cite{kotov2012electron,drut2009graphene,gamayun2010gap,gamayun2009supercritical}. However, there are no experimental indications of a gap-opening in suspended graphene~\cite{elias2011dirac}, for which $\alpha\approx 2.2$, to within 0.1 meV of the Dirac point. 

Driven or non-equilibrium DMs offer a new platform for investigation of collective instabilities. Recent optical pump-probe experiments on graphene have shown that the distribution of photoexcited carriers is highly non-thermal and can be effectively described by two separate Fermi-Dirac distributions with distinct chemical potentials for electrons and holes for around 100 fs after the excitation~\cite{george2008ultrafast, gilbertson2011tracing, li2012femtosecond, gierz2013graphene,gierz2015graphene}. Indications of population inversion have also been reported in 3DTIs with lifetimes of photoexcited carriers significantly larger than in graphene, from a few ps~\cite{zhu2015ultrafast} to $\mu$s for some samples~\cite{neupane2015gigantic}. 

Motivated by these experiments, we propose a scheme for generating transient many-body states in DMs by using external driving. By optically-pumping a DM, transient populations of electrons and holes are generated away from the nodal point allowing for a tunable enhancement of the effective coupling constant. In 2D, this tunability is unique to DMs and is not available in metals or semiconductors which possess a constant DOS at low energies. In such a system, electrons and holes at the two Fermi surfaces experience a mutual Coulomb attraction and can form electron-hole Cooper pairs, similar to Cooper pairs in the Bardeen-Cooper-Schrieffer (BCS) theory. At low temperatures such electron-hole pairs condense to form a superfluid phase known as the electron-hole BCS state, or an excitonic insulator~\cite{halperin1968possible, jerome1967excitonic}. This should be distinguished from a Bose-Einstein condensate (BEC) of excitons\cite{lozovik1976bec}, or bound states of a single electron-hole pair. Due to the non-equilibrium nature of electron and hole populations in pumped systems, we refer to this collective state as a transient excitonic condensate. 
 
Previous work has studied excitonic condensates in narrow-gap semiconductors~\cite{halperin1968possible, jerome1967excitonic}, electron-hole bilayers which are realizable in semiconductor heterostructures~\cite{lozovik1976superconductivity,zhu1995exciton} or graphene bilayers in the quantum Hall regime~\cite{eisenstein2004bose}, and in electronic systems under periodic driving~\cite{zhang2015floquet}. More recently, the possibility of inducing transient many-body states in semiconductors using optical driving has been studied theoretically~\cite{goldstein2015photoinduced}. Although transient excitonic condensates have not yet been observed experimentally in optically-pumped semiconductors, the signatures of preformed electron-hole pairs were measured in highly excited ZnO~\cite{versteegh2012pumpedZnO} which could be viewed as a precursor for the condensate.  

In this work, we propose to search for transient excitonic condensates in optically-pumped DMs. One signature of this state is the opening of gaps in the quasiparticle spectrum appearing at the two chemical potentials describing the electron and hole populations. In order to estimate the size of these excitonic gaps and critical temperatures for real materials, we use a simple model of a 2D DM with material specific parameters and with non-equilibrium electron and hole populations at different chemical potentials. Electron-electron interactions are treated at the mean-field level and we consider both the case of a simplified contact interaction and the screened Coulomb potential. 

We show that the critical temperature and the size of the excitonic gap is controlled by the interplay between the enhanced density of states at the non-equilibrium chemical potentials and metallic screening which becomes stronger with increasing the chemical potentials, the value of the coupling constant and the Dirac cone degeneracy. Based on this we derive a set of criteria to identify the best material candidates for observing the transient collective states.

Among the existing DMs, we predict the largest  effect, with the size of the gap of the order 10meV, in undoped suspended graphene in which optical pumping is realized selectively on a single valley, e.g. using circularly polarized light~\cite{liu2011chirality,hsu2015valley}. Such gap sizes are large enough to be detected by angle-resolved photoemmision spectroscopy (ARPES). We also find that DMs with a single non-degenerate Dirac cone and large coupling constants such as large-gap 3DTIs with small Dirac velocities and small dielectric constants, if realized, are the most promising candidates for observing the transient excitonic condensate. Our theoretical estimates indicate that gaps of the order of $100$~meV could be achieved in such materials. For all examples considered, we find critical temperatures that are several orders of magnitude larger than the estimated maximum critical temperature for excitonic condensate in double layer graphene~\cite{efetov}.

%We show that under certain circumstances the critical temperature and the size of the gap can be enhanced by increasing the values of the non-equilibrium chemical %potentials. In the case of graphene, we find that the gaps are large enough to be detected by angle-resolved photoemmision spectroscopy (ARPES). 
The rest of the paper is organized as follows. In Section~\ref{model}, we describe the details of our theoretical model. In Section~\ref{results}, we present the results of our calculations. In particular, we provide approximate expressions for the critical temperature of the transient excitonic condensate within a particular regime and we show the phase diagram for the excitonic condensate computed numerically. In Section~\ref{discuss}, we discuss our results and present the estimates of the excitonic gap and critical temperature for two important cases of 2D DMs studied in experiments, i.e. graphene and 3DTI surface states, as well as for a hypothetical 2D DM with parameters tuned in such a way as to reduce the screening effects. We also propose several schemes for experimentally detecting the transient excitonic instabilities in pumped DMs. Finally, in Section~\ref{concl} we offer concluding remarks.

\section{Theoretical Model}\label{model}
\begin{figure*}[ht!]
\centering
\includegraphics[width=0.8\linewidth,clip=true]{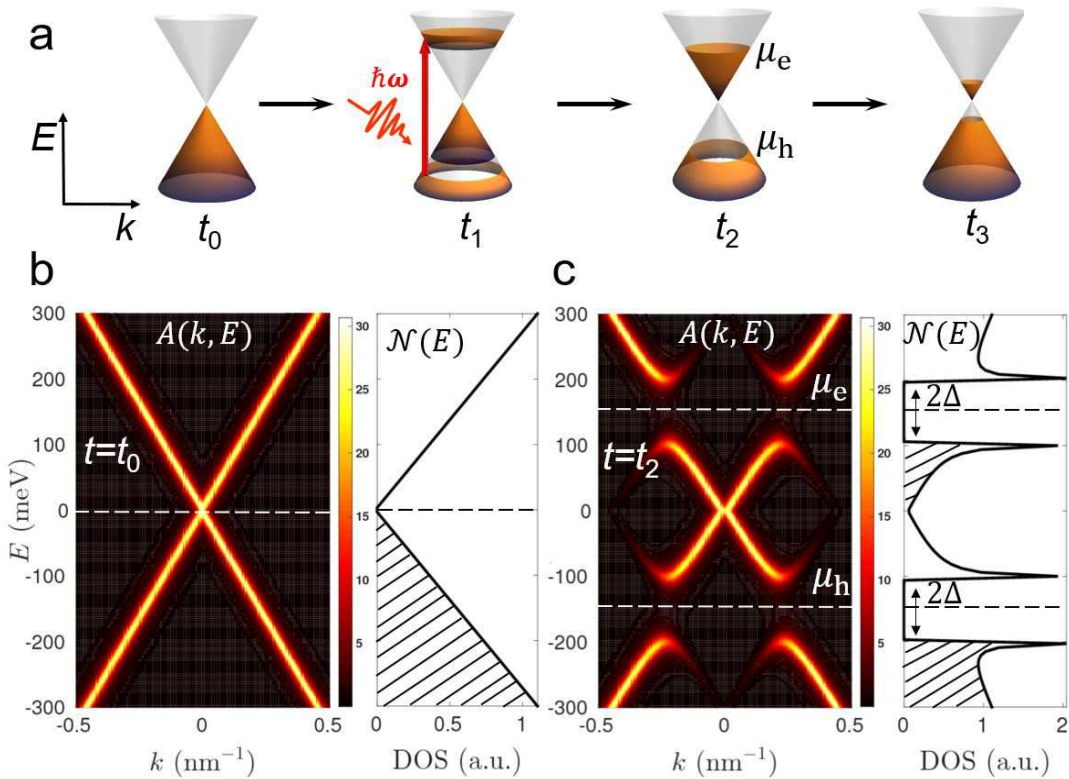}
\caption{Transient excitonic states in a pumped
DM. (a) Schematic of optical pumping considered in this work. Before the pump, at time $t=t_0$, the electrons exist in
equilibrium described by a single chemical potential ($\mu=0$
in the figure); at $t=t_1$ electrons are pumped from the valence
band to the conduction band far from equilibrium; shortly
after the excitation electrons and holes can be
described effectively by two distinct Fermi-Dirac distributions
with different chemical potentials, $\mu_{\mathrm{e}}$ and $\mu_{\mathrm{h}}$ ($t=t_2$); transient
populations eventually decay towards equilibrium ($t=t_3$). Empty states are shown in white, occupied states are shown in
yellow. (b) and (c) show the spectral
function, $A(k,E)$, and DOS, $\mathcal{N}(E)$, in equilibrium
and in a transient excitonic state, respectively, calculated using the contact interaction model. Shaded regions in the
DOS show occupied states.  For illustrative purposes, we use parameters for 
graphene on the substrate~\cite{kotov2012electron} with %$g=1$, 
$\alpha=1$, $v=6.58$~eV$\cdot\text{\AA}$, cutoff energy scale of $1$~eV, which corresponds to
momentum cutoff $\Lambda=1/6.58$~$\AA^{-1}$, $\mu_{-}=300$ meV, and $\mu_{+}=0$. Calculations were performed with the contact interaction model.
}
\label{fig1}
\end{figure*}

As a first step, we consider a simple model of a pumped 2D DM in which the electronic states take on a transient distribution governed by two chemical potentials, one for the electrons $\mu_\mathrm{e}$ and one for the holes $\mu_\mathrm{h}$ as shown in Fig.~\ref{fig1}(a). While the nature of photoexcited carriers in DMs at short time delays is an open issue, multiple optical pump-probe experiments have shown that such a transient population inversion could be achieved in graphene~\cite{george2008ultrafast, gilbertson2011tracing, li2012femtosecond} and 3DTIs~\cite{hajlaoui2014tuning, aguilar2015time, neupane2015gigantic, zhu2015ultrafast}. While there are other observations reporting a single hot Fermi-Dirac distribution in graphene tens of fs after the photoexcitation~\cite{johannsen2013graphene}, for the purposes of this paper, we assume that the population inversion in 2D DMs can be realized by optical pumping, as shown schematically in Fig.~\ref{fig1}(a).

Under realistic conditions, the transient state has a finite lifetime, limited to hundreds of femtoseconds (fs) in the case of graphene~\cite{george2008ultrafast, gilbertson2011tracing, li2012femtosecond,gierz2013graphene,gierz2015graphene} but extending to at least picoseconds (ps) in 3DTIs~\cite{zhu2015ultrafast}. Moreover, a recent time-resolved ARPES study by Neupane \textit{et al.}~\cite{neupane2015gigantic} revealed long-lived transient 3DTI surface states with lifetimes of a few microseconds ($\mu$s). Although the nature of such gigantic lifetimes is not entirely understood, this is a strong indication that long-lived (quasiequilibrium) excited Dirac states can be achieved in these systems.

We begin our analysis by assuming the lifetime of the transient populations of electrons and holes in pumped 2D DMs is sufficiently long that the system can be considered to be in quasiequilibrium. In this case, the system is described by the Hamiltonian $H=H_\mathrm{e}+H_\mathrm{h}+H_\mathrm{int}$ where
\begin{equation}
\begin{aligned}
H_{\mathrm{e/h}}&=\sum_{\textbf{k}} \xi^{(\mathrm{e/h})}_{\textbf{k}} \psi^\dagger_{\mathrm{e/h},\textbf{k}}\psi_{\mathrm{e/h},\textbf{k}} \\
H_{\mathrm{int}}&=\sum_{\textbf{q},\textbf{k},\textbf{k}'} V_{\textbf{q}} \psi^{\dagger}_{\mathrm{e},\textbf{k}+\textbf{q}}\psi^{\dagger}_{\mathrm{h},\textbf{k}^\prime-\textbf{q}}\psi_{\mathrm{h},\textbf{k}'}\psi_{\mathrm{e},\textbf{k}}.
\end{aligned}
\label{eq:H}
\end{equation}
Here $\xi^{(\mathrm{e})}_{\textbf{k}}=vk-\mu_\mathrm{\mathrm{e}}$ ($\xi^{(\mathrm{h})}_{\textbf{k}}=-vk-\mu_\mathrm{h}$) is the electron (hole) dispersion measured from the electron (hole) chemical potential, $\mu_\mathrm{\mathrm{e}}$ ($\mu_\mathrm{\mathrm{h}}$), where $v$ is the velocity of the Dirac states (in this work we set $\hbar\equiv 1$);
$\psi^\dagger_{\tau,\textbf{k}}$ ($\psi_{\tau,\textbf{k}}$) creates (annihilates) a Dirac state in band $\tau=\{\mathrm{e,h}\}$ with momentum $\textbf{k}$. Note that we consider a spinless model of a DM. $V_\textbf{q}=2\pi v\alpha/(q+\kappa)$ is the screened Coulomb potential modeled using the Thomas-Fermi theory (see derivation below), where $\alpha=e^2/\varepsilon v$ is the dimensionless coupling constant in the DM, $\varepsilon$ is the effective  dielectric constant and $\kappa$ is the screening wave vector~\cite{dassarma2011graphene, kotov2012electron}.

The order parameter, or gap, for an excitonic condensate in this system is given by $\Delta_{\textbf{k}}=\sum_{\textbf{k}'}V_{\textbf{k}-\textbf{k}'}\langle \psi_{\mathrm{e},-\textbf{k}'}\psi^\dagger_{\mathrm{h},\textbf{k}'} \rangle$. Combining this definition of the gap with the Hamiltonian in Eq.~(\ref{eq:H}), one can show that the gap equation is given by
\begin{equation}
\Delta_{\textbf{k}}=\frac{1}{2}\int \frac{d\textbf{k}'}{(2\pi)^2} V_{\textbf{k}-\textbf{k}'} \Delta_{\textbf{k}'}\frac{n_\mathrm{F}\left(\omega_{+}(\textbf{k}')\right)-n_\mathrm{F}\left(\omega_{-}(\textbf{k}')\right)}{\sqrt{(v|\textbf{k}'|-\mu_{-})^2+\Delta^2_{\textbf{k}'}}},
\label{eq2:gap}
\end{equation}
where $\omega_{\pm}(\textbf{k})=-\mu_{+}\pm\sqrt{(v|\textbf{k}|-\mu_{-})^2+\tfrac{1}{4}\Delta^2_{\textbf{k}}}$, $\mu_{\pm}\equiv (\mu_e\pm\mu_h)/2$, $n_\mathrm{F}(\omega)=1/(e^{\omega/T}+1)$ is the Fermi-Dirac distribution, and $T$ is the temperature (assumed to be identical for both photoexcited electrons and holes). The order parameter defined in Eq.~(\ref{eq2:gap}) represents the pairing between electrons and holes in a single non-degenerate Dirac cone and is assumed to be unaffected by the degeneracy of the Dirac states, which can be different from $1$, for instance, in graphene. However, the degeneracy will strongly affect the screening and will be taken into account~\cite{efetov}. By solving for the gap self-consistently we can study the conditions under which the quasiequilibrium Dirac states will condense to form excitonic gaps in the spectrum~\cite{halperin1968possible}. The gaps that open up at the electron and hole chemical potentials offer a signature of the transient excitonic condensate that can be probed by spectroscopic techniques. In Fig.~\ref{fig1}(b) we plot the spectral function, $A(k,E)$, and DOS, $\mathcal{N}(E)$, for a DM in equilibrium while in Fig.~\ref{fig1}(c) we show the corresponding plots for a DM with dynamically-generated excitonic gaps. These spectroscopic features in $A(k,E)$ and $\mathcal{N}(E)$ can be accessed experimentally using ARPES and scanning tunneling microscopy (STM) respectively.

The last term in Eq.~(\ref{eq:H}) describes the interband Coulomb interaction which is repulsive for pairs of electrons (or pairs of holes) but attractive  between electrons and holes. In the static limit, the screened Coulomb potential is given by~\cite{haug2004quantum} 
\begin{equation}
V_{\textbf{q}}=\frac{v_{\textbf{q}}}{\varepsilon(\textbf{q})},
\label{eq:coulomb_screened}
\end{equation} 
where 
\begin{equation}
v_{\textbf{q}}=\frac{2\pi e^2}{\varepsilon q}
\label{eq:coulomb_bare}
\end{equation} 
is the bare Coulomb potential in two-dimensional momentum space which can be obtained by Fourier transforming the long-range real-space potential $v(r)=e^2/\varepsilon r$, where $\varepsilon$ is the dielectric constant of the material and $e$ is the electron charge. In the random phase approximation, the $\mathbf{q}$-dependent dielectric function $\varepsilon(\textbf{q})$ is given by~\cite{haug2004quantum}
\begin{equation}
\varepsilon(\textbf{q})=1+\frac{\kappa}{q},
\label{eq:diel_func}
\end{equation} 
where $\kappa$ is the screening vector, or the inverse screening length of the combined system of electrons and holes. In 2D, the electron/hole screening wavenumber is given by 
\begin{equation}
\kappa_{i}=\frac{2\pi e^2}{\varepsilon}\frac{\partial n_{i}}{\partial \mu_{i}},
\label{eq:screen_wavevector_eh}
\end{equation} 
where $n_i$, $i=\{\text{e},\text{h}\}$, is the density of electrons or holes. One can see that $\kappa_{i}\propto \mathcal{N}(\mu_i)$,  where $\mathcal{N}(\mu_i)$ is the DOS at 
the electron or hole chemical potential. (Note that in 3D, $\kappa_i\propto\sqrt{\mathcal{N}(\mu_i)}$)~\cite{haug2004quantum}. At $T=0$, Eq.~(\ref{eq:screen_wavevector_eh}) is referred 
to as the Thomas-Fermi screening wavevector. In this case, the density is related to the Fermi wavevector, $k^i_\mathrm{F}$, as $k^i_\mathrm{F}=\sqrt{4\pi n_{i}/g}$, 
where $g$ is the Dirac cone degeneracy.
For a Dirac spectrum, $k^i_\mathrm{F}=|\mu_i|/v$ and therefore $\kappa_i=\tfrac{2\pi e^2}{\varepsilon}\tfrac{\partial n_i}{\partial \mu_i}=\tfrac{g e^2 k^i_\mathrm{F}}{\varepsilon v}=g\alpha k^i_\mathrm{F}$. 
 %, where $\alpha=e^2/\varepsilon v$ is the dimensionless coupling constant. 
In 2D the screening wavenumber of the electron-hole plasma is given by $\kappa=\kappa_\mathrm{e}+\kappa_\mathrm{h}$~\cite{klingshirn2005optics}. 
For equal chemical potentials ($\mu_{+}=0$), $\kappa_\mathrm{e}=\kappa_\mathrm{h}$ and $\kappa=2\kappa_{\mathrm{e}}$. Hence, $\kappa\propto g\alpha\mu_{-}$, 
and the screening  becomes stronger for  larger chemical potentials, larger $\alpha$ and larger $g$.

Substituting expressions for the bare Coulomb potential $v_{\mathbf{q}}$ [Eq.~(\ref{eq:coulomb_bare})] and the dielectric function $\varepsilon(\mathbf{q})$ [Eq.~(\ref{eq:diel_func})] into Eq.~(\ref{eq:coulomb_screened}), the screened Coulomb potential can be re-written as 
\begin{equation}
V_{\mathbf{q}}=\frac{2\pi e^2}{\varepsilon(q+\kappa)}=\frac{2\pi\alpha v}{q+\kappa}
\label{eq:coulomb_screened_final}.
\end{equation}
In the following section we will solve the self-consistent gap equation [Eq.~(\ref{eq2:gap})] first using a simplified interaction and then the screened Coulomb potential defined in Eq.~(\ref{eq:coulomb_screened_final}). 

\section{Results}\label{results}
\subsection{Analytical results for contact interaction}\label{results_contact}
We can gain some insight into the behavior of these systems by analyzing the limiting case in which the screening of the Coulomb interaction is so severe that the interaction potential becomes a contact interaction in position space or, equivalently, a constant in momentum space $V_\textbf{q}=V_0=2\pi v \alpha/\kappa$. We expect the analysis with the contact interaction to agree quantitatively with the screened Coulomb interaction when $\kappa>>\Lambda$ where $\Lambda$ is the momentum cutoff for the Dirac model in Eq.~(\ref{eq:H}).

Assuming the interaction potential in Eq.~(\ref{eq:H}) is given by $V_0$, we can see that the right-hand side of the gap equation, Eq.~(\ref{eq2:gap}), becomes independent of $\textbf{k}$; therefore, $\Delta(\textbf{k})=\Delta_0$ is constant in momentum. In this case, we can perform the angular integration analytically and find approximate expressions for the radial integral in terms of the momentum cutoff (see Appendix \ref{contact_analysis}). We can then solve the resulting equation for the critical temperature $T_\mathrm{c}$ as a function of the average and difference of the two chemical potentials, $\mu_{\pm}$
\begin{equation}
\begin{aligned}
T_\mathrm{c}(\mu_{+},\mu_{-})&\approx \overline{T}_\mathrm{c}(\mu_{-})\left(1 - C_{+} \frac{\mu_{+}^2}{4\overline{T}_\mathrm{c}(\mu_{-})^2} \right); \\
\overline{T}_\mathrm{c}(\mu_{-})&=C_{-} \sqrt{\mu_{-}(v\Lambda-\mu_{-})}\exp\left( \frac{v(\Lambda-2\kappa\alpha^{-1})}{2\mu_{-}}\right)
\end{aligned}
\label{eq:T_c}
\end{equation}
where $C_{+}=\gamma+\ln{\frac{4}{\pi}}-1/3$, $C_{-}=\frac{2}{\pi}e^{\gamma-1}$, where $\gamma\approx 0.5772$ is the Euler-Mascheroni constant.

From Eq.~(\ref{eq:T_c}) we observe several important trends. One crucial feature is that, as expected, $T_\mathrm{c}$ increases with increasing $\mu_{-}$. In fact, for perfectly matched chemical potentials ($\mu_{+}=0$) and when $2\kappa/\alpha\leq\Lambda$ any finite value of $\mu_{-}$ leads to the formation of a gapped state as shown in Fig.~\ref{fig1}(c). This confirms the intuitive argument that an excitonic condensate is expected to form in such a quasiequilibrium state as a consequence of the enhanced DOS away from the Dirac node.

Another key feature is that away from perfect matching ($\mu_{+}\neq0$) $T_\mathrm{c}$ decreases. To leading order in $\mu_{+}/2T$ the decrease is quadratic and for $\mu_{+}/2T>>1$, $T_\mathrm{c}$ vanishes (see Appendix \ref{contact_analysis}). Therefore, we expect that excitonic gapped states should be most easily realized in systems with little screening, strong coupling, and matched chemical potentials, as in the case of undoped suspended graphene.

\subsection{Numerical results for screened Coulomb interaction}\label{results_coulomb}
To provide quantitative estimates of $T_\mathrm{c}$ we consider the more realistic case of a screened Coulomb potential given by Eq.~(\ref{eq:coulomb_screened_final}). 
Unlike the contact interaction, the potential in Eq.~(\ref{eq:coulomb_screened_final}) accounts for both the long-range nature of the electron-electron interaction and the 
metallic screening which becomes important when the chemical potential is shifted away from the Dirac node. In the case of the Coulomb potential, the gap, $\Delta_\mathbf{k}$, 
is momentum-dependent with a maximum value at $|\mathbf{k}|=k_\text{F}$. We proceed by solving the self-consistent gap equation, Eq.~(\ref{eq2:gap}), numerically, assuming an isotropic gap.

\begin{figure}[ht!]
\centering
\includegraphics[width=0.98\linewidth,clip=true]{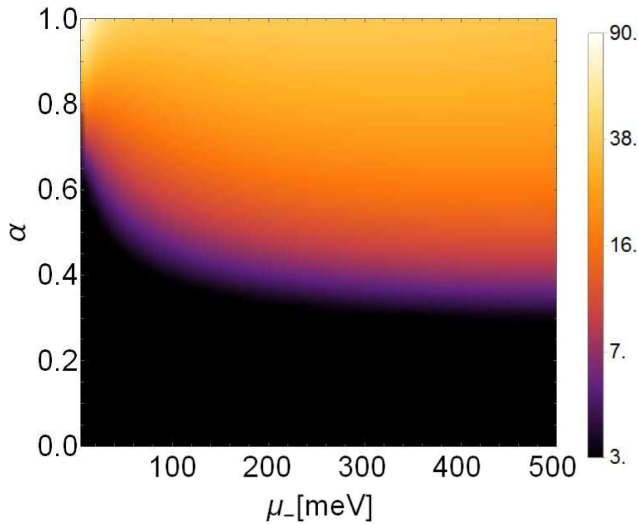}
\caption{Tunability of the criticial coupling. 
The maximum value of the gap evaluated at $k=k_\mathrm{F}$ as a function
of   $\mu_{-}$ and $\alpha$ for matched chemical potentials ($\mu_{+}=0$), $T=0$ and $g=1$. 
Other parameters are the same as in Fig.~\ref{fig1}. The color shade represents the maximum value of the gap  in meV's and is logarithmically scaled for better contrast.}
\label{fig2}
\end{figure}

In Fig~\ref{fig2} we plot the phase diagram of the transient excitonic state in the $\mu_{-}-\alpha$ plane. As in the case of the contact 
interaction, the critical coupling, $\alpha_\mathrm{c}$, defined for a given $\mu_{-}$ as the value of $\alpha$ for which $\Delta_\mathbf{k}$ is different from zero, 
%\footnote{In our numerical simulations we use a condition $\Delta_{k_\mathrm{F}}\ge\delta$, where $\delta$ is a small number ($\delta=10^{-6}$ in units of energy).} 
is dramatically reduced in the quasiequilibrium state, $\mu_{-}\neq 0$. This phase diagram confirms that the qualitative predictions made by Eq.~(\ref{eq:T_c}) hold 
for the case of the full Coulomb interaction and lends further support to the premise that excitonic instabilities can be dynamically-induced in DMs.

\begin{figure*}[ht!]
\centering
\includegraphics[width=0.98\linewidth,clip=true]{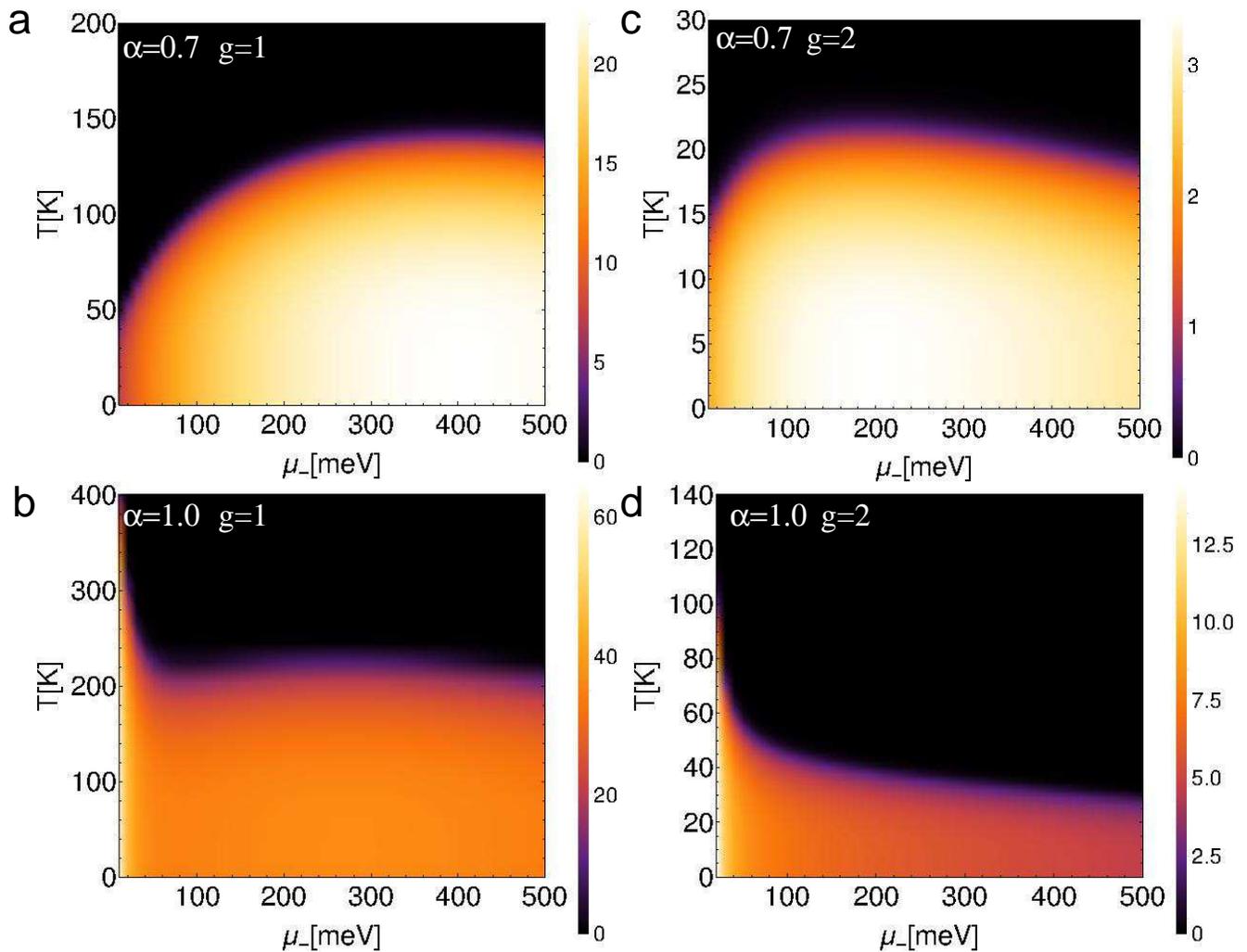}
\caption{Phase diagram of the transient excitonic condensate. 
The maximum value of the gap evaluated at $k=k_\mathrm{F}$ as a function 
of $\mu_{-}$ and $T$ for matched chemical potentials ($\mu_{+}=0$) computed numerically by solving Eq.~(\ref{eq2:gap}) self-consistently using the Thomas-Fermi potential shown in Eq.~(\ref{eq:coulomb_screened_final}) for (a, c) $\alpha=0.7$ and (b, d) $\alpha=1.0$. Left column is for $g=1$, right column is for $g=2$.  Other parameters are the same as in Fig.~\ref{fig1}. The color shade represents the maximum value of the gap in meV's. Note that we exclude the region around $\mu_{-}=0$ since it is not properly described in the Thomas-Fermi model.}
\label{fig3}
\end{figure*}

%T-mu phase diagram for two values of alpha, below alpha_c_0 and above alpha_c_0
In Fig.~\ref{fig3} we show the maximum of the gap calculated with the full Coulomb potential plotted in the $T-\mu_{-}$ plane for two regimes: $\alpha<\alpha_\mathrm{c}^0$, in (a) and (c); 
and $\alpha>\alpha_\mathrm{c}^0$, in (b) and (d), where $\alpha_\mathrm{c}^0$ is the equilibrium critical coupling. (Our 
mean-field model gives $\alpha_\mathrm{c}^0\approx 1.0$~\cite{numerical_note}.) In each regime we consider two cases for the degeneracy, $g=1$ and $g=2$. 
 Unlike in the case of the contact-interaction model, where the gap and $T_\mathrm{c}$ always exhibit 
an increase with $\mu_{-}$ near the Dirac point (see Fig.~B\ref{fig:T_c_contact} in Appendix~\ref{contact_analysis}), in the case of the screened Coulomb potential the phase diagram 
for the order parameter exhibits a more complex interplay between screening, which becomes stronger with increasing $\alpha$, $\mu_{-}$ and $g$, the value of the DOS at a given $\mu_{-}$ 
and the value of $\alpha$. 

In the case $\alpha<\alpha_\mathrm{c}^0$ and $g=1$, Fig.~\ref{fig3}(a), the gap is vanishingly small around $\mu_{-}\approx 0$ and increases with increasing $\mu_{-}$, similar to the 
result of the contact-interaction model, until it reaches a maximum. For large $\mu_{-}$ the screening becomes strong leading to a decrease of the gap and $T_\mathrm{c}$. Similar behavior 
is observed for the same value of $\alpha$ and $g=2$ as shown in Fig.~\ref{fig3}(c). However, the downturn in $T_\mathrm{c}$ occurs at  smaller values of $\mu_{-}$ due to larger screening.
 
In the case $\alpha\gtrsim\alpha_{\mathrm{c}}^0$ and $g=1$, Fig.~\ref{fig3}(b), the gap is different from zero already at 
equilibrium and is enhanced at small $\mu_{-}$ due to the diverging Thomas-Fermi screening length which leads to an essentially unscreened Coulomb potential. 
As $\mu_{-}$ increases the screening becomes stronger and $T_\mathrm{c}$ decreases rapidly. This is followed by an upturn in the size of the gap and $T_\mathrm{c}$ 
at $\mu_{-}\approx 50$~meV due to the enhanced DOS. Finally, the gap starts to decrease as the screening becomes dominant at 
large $\mu_{-}$. For $\alpha\gtrsim\alpha_{\mathrm{c}}^0$ and $g=2$,  Fig.~\ref{fig3}(d), the behavior at small $\mu_{-}$ is similar to the $g=1$ case. However, 
at large $\mu_{-}$ the gap and $T_c$ decrease monotonically due to severe screening, making this case less favorable compared to the $g=1$ case.

\begin{figure}[ht!]
\centering
\includegraphics[width=\linewidth,clip=true]{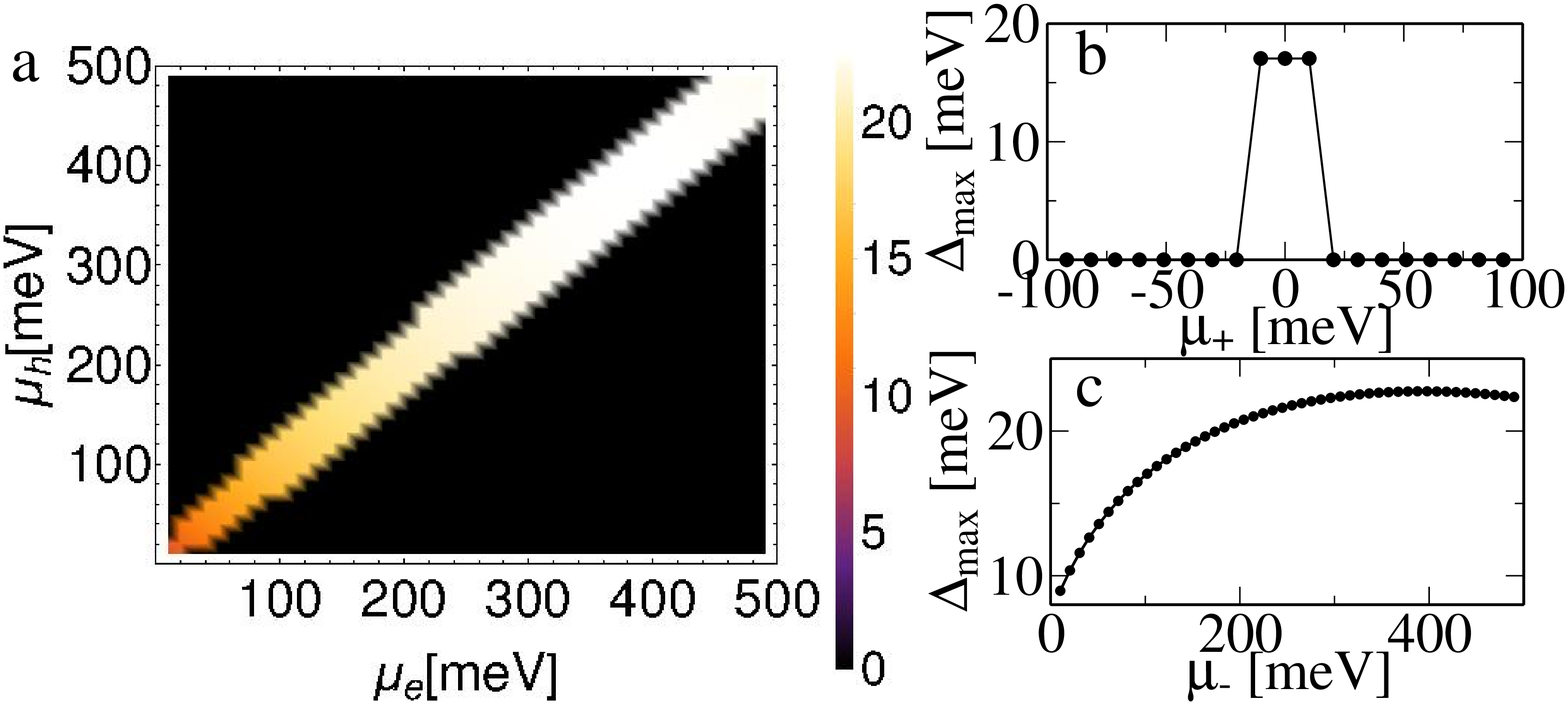}
\caption{Stability diagram of the transient excitonic condensate as a function of chemical potential mismatch. 
The maximum value of the order parameter $\Delta_\textbf{k}$ evaluated at $T=0$ plotted as a function of (a) 
electron and hole chemical potentials, (b) the mismatch $\mu_{+}$ for a fixed average chemical potential $\mu_{-}=0.1$~meV, and (c) 
the average chemical potential $\mu_{-}$ for zero mismatch $\mu_{+}=0$. We used $g=1$ and $\alpha=0.7$.}
\label{fig_unb}
\end{figure}

Figure~\ref{fig_unb} shows the maximum of the gap calculated with the full Coulomb potential as a function of the mismatch between the electron and hole chemical potentials at $T=0$. 
A domain of stability of the excitonic condensate exists in the region of the parameter space where the gap is different from zero, Fig.~\ref{fig_unb}(a). In Fig.~\ref{fig_unb}(b) 
it is clear that the gap reaches its maximum at $\mu_{+}=0$ and quickly vanishes away from $\mu_{+}=0$, in agreement with the prediction of the contact interaction model 
[see Eq.~(\ref{Tc_unb_AppB}) in Appendix~\ref{contact_analysis}]. Furthermore, we confirm that for this value of $g$ and $\alpha$ ($g=1$ and $\alpha=0.7$ in Fig.~\ref{fig_unb}) the magnitude of 
the gap increases with increasing $\mu_{-}$, Fig.~\ref{fig_unb}(c).  

\section{Discussion and Experimental Feasibility}\label{discuss}
In order to test the feasibility of our proposal for achieving a transient excitonic-insulator state in a pumped DM, we provide estimates of the critical temperature and the gap for different materials in typical experimental setups. Table~\ref{tab:values} summarizes the results for graphene on a substrate, free-standing graphene, and 3DTIs such as binary bismuth chalcogenides and related materials. As input parameters for our model we use the material-specific coupling constant and  the typical pump energy (see Appendix~\ref{params} for discussion of the choice of material parameters). In addition, we list the expected lifetimes ($\tau$) of the transient excitonic state inferred from experimental data for existing DMs. An important parameter is the Dirac cone degeneracy. For 3DTI, we take $g=1$. For graphene, we consider $g=4$ and $g=2$, corresponding to conventional pumping with linearly polarized light and valley-selective pumping, respectively, with the latter being the most favorable situation. We also present  the estimates for a hypothetical DM with parameters similar to those of graphene and degeneracy $g=1$, which could be realized in large-gap 3DTIs with a single Dirac cone and a large $\alpha$.

\begin{table*}[ht]
\caption{Critical temperatures and excitonic gaps in pumped DMs. Dimensionless coupling constant $\alpha$, average chemical potential $\mu_{-}$ and lifetime of the transient excitonic state $\tau$ for graphene and 3DTIs were estimated based on existing literature (see Appendix~\ref{params} for details). Critical temperature $T_\mathrm{c}$ and maximum value of the excitonic gap $\Delta_\mathrm{max}$ at $T=0$ for given material parameters were calculated using the numerical model with screened Coulomb interaction. For the case of graphene we consider both the one valley ($g=2$) and two valley ($g=4$) cases. The last row corresponds to a hypothetical DM with parameters similar to graphene and $g=1$.}
\label{tab1}
\begin{tabular}{r||l|l|l|c|r}
Material &  $\alpha$ &  $\mu_{-}$ (meV) & $\tau$ (ps) &  $T_\mathrm{c}$ (K) & $\Delta_\mathrm{max}$ (meV)\\
\hline
\hline
graphene (substrate) &  $0.4-1.0$ & $500$ & $0.1$ & \begin{tabular}{rl} $1.0$ & for $g=4$ \\  $6-35$ & for $g=2$ \end{tabular}  
 & \begin{tabular}{rl} $0.1$ & for $g=4$ \\  $1-5$ & for $g=2$ \end{tabular} \\
\hline
 graphene (suspended)         & $2.2$ & $500$ & 0.1 & \begin{tabular}{rl} $2$ & for $g=4$ \\  $70$ & for $g=2$ \end{tabular} 
 & \begin{tabular}{rl} $0.3$ & for $g=4$ \\  $10$ & for $g=2$ \end{tabular}   \\ % free-standing graphene
\hline
3DTI    & $0.1-1.0$ & $100$ & $1-10^{6}$ & $0-30$ & $0-3$  \\
\hline
DM with g=1 &  $0.4-2.2$ & $500$ & - & $50-500$ & $10-100$ \\
\hline			
%DM with g=2 & $0.4-2.2$ & 500 & - & $6-70$ & $1-10$ \\
\end{tabular}
\label{tab:values}
\end{table*}
               %\alpha         %mu      %T_c              $\Delta
%DM with g=4 & $0.4-1.0-2.2$ & $500$ & $0.75-1.25-2$ & $0.1-0.2-0.3$ \\
%DM with g=2 & $0.4-1.0-2.2$ & $500$ & $6-35-70$ & $1-5-10$ \\
%DM with g=1 & $0.4-1.0-2.2$ & $500$ & $50-500$ & $10-100$ \\ 

To relate the values of the non-equilibrium chemical potentials to the pump energy in the case of graphene, we start by noting that $\mu_\mathrm{e/h}=v\sqrt{\pi n_\mathrm{e/h}}$, where $v$ is the system-specific Fermi velocity and $n_\mathrm{e/h}$ is the density of photoexcited carriers. The density of photoexcited carriers can be estimated using $\hbar\omega_\mathrm{pump}n_\mathrm{e/h}\approx \Phi A_0$~\cite{li2012femtosecond}, where $\Phi$ is the pump fluence and $A_0$ is the absorption coefficient of graphene. We estimate that $\mu_{-}=500$~meV will be achieved with carrier densities $n_\mathrm{e/h}\approx 10^{13}$~cm$^{-2}$, which corresponds to a fluence of $\Phi\approx 100$~$\mu$~Jcm$^{-2}$ for $A_0\approx 0.02$ and $\hbar\omega_\mathrm{pump}\approx 1$~eV. These values are in agreement with experimental estimates~\cite{brida2013ultrafast}.

In the case of suspended graphene and single-valley pumping ($g=2$), for an average chemical potential $\mu_{-}=500$~meV, assuming balanced chemical potentials ($\mu_{+}=0$), we predict the maximum size of the gap to be $\Delta_\mathrm{max}\approx 10$~meV for temperatures up to $\approx 100$~K. Such gap sizes are within the energy resolution of ARPES. Time-resolved optical conductivity measurements~\cite{gilbertson2011tracing} may provide an alternative probe for verifying the presence or absence of the gaps in the spectrum. Additionally, the excitonic gap opening should result in an enhanced photoluminescence due to the recombination of electron-hole pairs~\cite{versteegh2012pumpedZnO}.

It should be noted that the peak temperature of the inverted carrier distributions inferred from pump-probe experiments can be as large as a few thousand Kelvin~\cite{gierz2013graphene, li2012femtosecond}. While this is above the predicted $T_\mathrm{c}$ for the typical DMs in Table~\ref{tab1}, an increased lifetime of the transient state should lead to lower local electronic temperatures as the carriers have more time to cool down. 
Importantly, the predicted critical temperatures for pumped graphene are several orders of magnitude larger than the estimated maximum critical temperature of excitonic condensation in double layer graphene in the static regime ($T_\mathrm{c}^{\mathrm{max}}\lesssim 1$mK)~\cite{efetov}. Our calculations predict that even larger $T_\mathrm{c}$ of hundreds of K could be in principle achieved in future DMs with carefully designed material properties.
 
In Table~\ref{tab:values}, we see that $\alpha$ in 3DTIs appears to be an order of magnitude smaller than that found in graphene due to the large dielectric constants of 3DTIs~\cite{kim2012ticonduction}. Using values of $\varepsilon\approx 100$ and $v\approx 5\times 10^{5}$~m/s we find that $\alpha\approx 0.1$ and that the predicted excitonic gap is $<1$~meV, which is below the current resolution of typical ARPES experiments. This drawback could be overcome in a 3DTI with a larger effective coupling constant [see Table~\ref{tab:values}], which would correspond to either a smaller dielectric constant or Fermi velocity~\cite{triola2015many}. Smaller Fermi velocities can be found in anisotropic Dirac cones on various crystal facets of topological insulators~\cite{pertsova2016quantum}. Recently, tilted Dirac cones with Fermi velocities of the order of $10^4$~m/s leading to large $\alpha$ have been found in some quasi-two-dimensional organic conductors~\cite{hirata2016organic,hirata2016organic2}.

In the above analysis of the transient inverted population quasiequilibrium was assumed; however, due to multiple scattering processes, the non-equilibrium inverted carrier distribution decays toward equilibrium and, as a result, the excitonic states in a pumped DM acquire a finite lifetime. We have analyzed the relaxation of the order parameter using a dynamical approach based on semiconductor Bloch equations~\cite{winzer2013microscopic, stroucken2011optical} applied to a pumped DM, where both intraband relaxation and interband scattering (recombination) is taken into account (see Appendix~\ref{time_evolution} for details). The result is that the characteristic timescale over which the population inversion is sustained provides an estimate of the lifetime of the transient excitonic gapped state.

So far the lifetime of the inverted carrier distribution observed in graphene has been limited to hundreds of fs~\cite{gierz2013graphene} after which a single equilibrium Fermi-Dirac distribution is reestablished via inverse Auger scattering (recombination) and electron-phonon scattering~\cite{gierz2015graphene}. However, these relatively short lifetimes were obtained for hole-doped graphene with the equilibrium chemical potential lying a few hundred meV below the Dirac node, as is typical for graphene on a substrate. One might expect more favorable conditions in undoped graphene, in which recombination of carriers is suppressed due to reduced phase space near the Dirac node. Additionally, one possible way to increase the lifetime of the inverted population is to use continuous pumping where electrons are constantly injected into the empty states above the Dirac node. However, such a scheme might result in high local electronic temperatures which will inhibit the formation of the excitonic condensate. In 3DTIs the reported lifetimes of the population inversion are much longer, e.g. of the order of a few ps~\cite{zhu2015ultrafast} and under certain conditions even exceeding $4\mu$s~\cite{neupane2015gigantic}.  

In general, in order for the transient excitonic gaps to be observable in experiments, the timescale of formation of these collective states, $\tau_\mathrm{ex}=\hbar/\Delta_\mathrm{max}$, should be small compared to the timescale $\tau_\mathrm{el}$ on which the inverted population is observed, $\tau_\mathrm{ex}<<\tau_\mathrm{el}$. For $\Delta_\mathrm{max}\approx 10$~meV, $\tau_\mathrm{ex}\approx 60$~fs, allowing for the observation of these gaps in graphene where $\tau_\mathrm{el}$ is on the order of $100-200$~fs. For smaller gaps, larger lifetimes of inverted population are necessary which could be realized in future experiments on graphene or 3DTIs.

While our analysis focused on 2D DMs, like graphene or 3DTI surface states, our theory also has implications for 3D DMs such as Dirac~\cite{neupane2014dirac} and Weyl~\cite{huang2015weyl,xu2015weyl} semimetals. In a 3D DM, the DOS is quadratic in energy, $\mathcal{N}\propto E^{2}$, in contrast to the linear dependence in 2D, hence the effective coupling can be made even larger in these materials. However, since screening could also be much stronger in 3D a more detailed analysis would be necessary to make quantitative predictions for these materials. It should be noted that the large valley degeneracy found in some 3D Dirac systems ($g=24$ in TaAs Weyl semimetal~\cite{huang2015weyl}) could be detrimental for the effects discussed in this paper. Therefore, as in the case of 2D DMs, materials with smaller degeneracy will be the most promising candidates.

\section{Conslusions}\label{concl}
In conclusion, we have shown that the energy-dependence of the DOS in 2D Dirac materials allows for a tunable enhancement of the strength of the Coulomb interaction relative to the values accessible in equilibrium. We have demonstrated that this tunability allows for the generation of transient excitonic states in optically-pumped 2D Dirac materials leading to the formation of gaps in the quasiparticle spectrum away from the Dirac node. Our estimates indicate that these dynamically-induced gaps can be as large as $10$~meV in the case of graphene and a few meV for 3DTI. With these results we have proposed an experimental scheme in which these excitonic gaps could be detected via pump-probe spectroscopy on undoped graphene and 3DTIs. Finally, we have provided guidelines for the search for novel Dirac materials with improved properties in which larger gaps and critical temperatures could be observed. 

\section*{Acknowledgement}
This work was supported by ERC-DM-32031, KAW, CQM. Work at LANL was supported by USDOE BES E3B7. We acknowledge support 
from Dr. Max R{\"{o}}ssler, the Walter Haefner Foundation and the ETH Zurich Foundation. We wish to thank David Abergel, Yaron Kedem, 
Rohit Prasankumar, Sergey Pershoguba, Antoinette Taylor, D. Yarotski, N. Plumb, Marijn Versteegh and Vladimir Juri\v{c}i\'{c} for useful discussions.

%\textbf{Author Contributions} A. P. and C. T. performed the calculations, discussed the results and wrote the MS. R. M. performed the calculations and discussed the results. A.V. B. proposed and supervised the projects, discussed the results and edited the MS. 
%\textbf{Competing financial interests} The authors declare that they have no
%competing financial interests.
%\textbf{Correspondence} Correspondence and requests for materials
%should be addressed to ~(email: balatsky@kth.se).

\appendix

\section{Tunability of the effective coupling constant}\label{params}
A great deal of work has gone into the study of the phase diagram of graphene with respect to the dimensionless coupling constant 
$\alpha=e^2/\varepsilon v$~\cite{dahal2010charge, kotov2012electron, dahal2006absence, drut2009graphene, khveshchenko2009massive, ryu2009masses, 
gorbar2002magnetic, gamayun2010gap, gamayun2009supercritical, triola2015many} where $e$ is the charge of the electron, $\varepsilon$ is the system-specific 
dielectric constant, and $v$ is the velocity of the Dirac electrons. These results suggest the existence of a critical value $\alpha_\mathrm{c}$ such that if 
$\alpha<\alpha_\mathrm{c}$ the spectrum remains gapless while if $\alpha\geq\alpha_\mathrm{c}$ the system flows toward the strong coupling regime~\cite{kotov2012electron}. 
In the strong coupling regime pairs of electrons and holes bind to form excitons which condense giving rise to a gap in the quasiparticle spectrum of the ground state. Thus 
far, perturbative and numerical results for graphene suggest the critical value is $\alpha_\mathrm{c}\approx 1$~\cite{kotov2012electron, drut2009graphene, gamayun2010gap, 
gamayun2009supercritical}. Much like previous work, our mean-field model of a two-dimensional (2D) Dirac material (DM) in equilibrium ($\mu=0$) gives a critical value 
$\alpha_\mathrm{c}\approx 1$. However, experiments involving suspended graphene, for which $\alpha\approx 2.2$, seem to indicate a gapless state to within 0.1 meV of the 
Dirac point~\cite{elias2011dirac} likely due to the logarithmic increase of the Fermi velocity close to the Dirac point\cite{kotov2012electron,elias2011dirac}. 

In a pumped DM with an inverted population there will be a finite density of both electrons and holes which will experience an attraction proportional to the density of 
states (DOS) of the two species times the strength of the coupling. Since the DOS is linear in energy this factor will be determined by the parameters 
$\mu_{\pm}\equiv (\mu_\text{e}\pm\mu_\text{h})/2$, while the coupling will be determined by the strength of the Coulomb interaction, Eq.~(\ref{eq:coulomb_screened}), 
which is controlled by $\alpha$ and screening effects. Therefore the effective interaction can be tuned either by directly modifying the dimensionless coupling constant 
$\alpha$ or by tuning the DOS. Table~\ref{alpha_numbers} contains the estimates of $\alpha$ for graphene and three-dimensional topological insulators (3DTIs) based on typical 
values of the dielectric constants and velocities found in the literature. 

\begin{table}[ht!]
\caption{Estimates of the dimensionless coupling constant in graphene (free-standing and on the substrate) and 3DTI for typical values of the dielectric constant 
and velocity of the Dirac states.}
\label{tabs1}
\begin{tabular}{l||l|l|c}
\hline
Material  & $\varepsilon$ & $v$ ($10^{6}$m/s) &  $\alpha$ \\
\hline
\hline
graphene (suspended) &  $1$ & $1.0$  & $2.2$ \\
\hline
graphene (substrate) &  $2-15$ & $1.0$  & $0.1-1.0$ \\
\hline
3DTI    & $30-50$ & $0.2-0.6$ & $0.1-0.4$ \\
\hline
\end{tabular}\label{alpha_numbers}
\end{table}

The velocity of the Dirac states in graphene is given by $v\approx1.0\times 10^{6}$m/s while in typical 3DTIs, such as binary bismuth chalcogenides and related materials, 
$v=2.0-6.0 \times 10^{5}$~m/s. Dirac states on the $(110)$ surface of Bi$_2$Se$_3$ have $v\approx 5\times 10^{5}$m/s. One can find smaller velocities on other crystal facets 
of 3DTIs. For example, the $(\bar{1}12)$ surface of Bi$_2$Se$_3$ hosts anisotropic (tilted) Dirac cones where the velocity in the vertical direction (along the direction of 
quintuple-layer growth) is $v\approx 2\times 10^{5}$m/s~\cite{pertsova2016quantum}.

In the case of graphene, the effective dielectric constant is taken to be $\varepsilon=(\varepsilon_\mathrm{sub}+\varepsilon_\mathrm{vac})/2$, where $\varepsilon_\mathrm{sub(vac)}$ 
is the dielectric constant of the substrate(vacuum). In the case of 3DTIs, $\varepsilon=(\varepsilon_\mathrm{TI}+\varepsilon_\mathrm{vac})/2$, where $\varepsilon_\mathrm{TI}$ is the 
dielectric constant of the bulk 3DTI. The effective dielectric constant depends strongly on the environment and also on the applied electric field~\cite{santos2013tmd}. 
Reported values of $\varepsilon$ in graphene on the substrate are in the range between $2$ to $16$, which gives $\alpha\in[0.1:1.0]$ (see Ref.~\onlinecite{santos2013tmd} and 
references therein). For two typical substrates such as SiC and SiO$_2$, $\alpha\approx 0.4$ and $\alpha\approx 0.8$, respectively, while for free-standing graphene 
(nominally $\varepsilon=1$), $\alpha\approx2.2$. 

In 3DTIs, the dielectric constant can be quite large e.g. $\varepsilon_\mathrm{TI}\approx 100$ ($113$ in Bi$_2$Se$_3$ and $75-290$ in Bi$_2$Te$_3$~\cite{richter1977dielectric}), 
which gives $\varepsilon\approx 50$. In some experiments $\varepsilon$ has been taken to be closer to $30$ due to heavy doping of the samples~\cite{Beidenkopf2011spacial}. 
For typical velocities in Table~A\ref{alpha_numbers}, this gives $\alpha\in[0.1:0.4]$. However, since $v$ can be made smaller in some cases~\cite{triola2015many} and $\varepsilon$ 
could be tuned, in principle, by gating in TI thin films, we investigate a larger range of $\alpha$, $\alpha\in[0.1:1.0]$, similar to the case of graphene. 

In a pumped DM with non-zero chemical potential of photoexcited electrons and holes, the DOS is determined by the values of the chemical potentials that can be achieved in experiment. 
In the case of graphene, we can relate the chemical potentials $\mu_\mathrm{e/h}$ to the number of photoexcited carriers $n_\mathrm{e/h}$ as $\mu_\mathrm{e/h}=v\sqrt{\pi n_\mathrm{e/h}}$. 
The number of carriers can be estimated using the properties of the pump pulse, i.e. for a pump fluence $\Phi$ and a pump 
energy $\hbar\omega_\mathrm{pump}$, $\hbar\omega_\mathrm{pump}n_\mathrm{e/h}\approx \Phi A_0$~\cite{li2012femtosecond}, where $A_0$ is the absorption coefficient of graphene. 
Taking $A_0=0.02$, $\hbar\omega_\mathrm{pump}=1$~eV and assuming a balanced distribution of photoexcited electrons and holes ($n_\mathrm{e}=n_\mathrm{h}\equiv n_\mathrm{ex}$), we estimate that in order to achieve a chemical potential $\mu_{\mathrm{e/h}}=\mu_{-}\approx 500$meV, one would need a carrier density $n_\mathrm{ex}\approx 10^{13}$~cm$^{-2}$ corresponding to a pump fluence $\Phi\approx 100$~$\mu$Jcm$^{-2}$. Such carrier densities and pump fluences can be achieved in present experiments~\cite{brida2013ultrafast}. Thus, we used the value $\mu_{-}=500$~meV for estimates of the excitonic gap and critical temperature in graphene. 

In 3DTIs the pump energy is typically large compared to the bulk bandgap, e.g. $\hbar\omega_\mathrm{pump}\approx 1.5$~eV while the bulk bandgap in chalcogenide 3DTIs is about $0.3$~eV. Therefore electrons are first excited into empty states in the bulk conduction band and then quickly populate the lower-energy Dirac states. As observed in recent experiments, %the net flow rate of photoexcited electrons from high to low energy decreased (the lifetime increases) 
the lifetime increases as the energy approaches the Dirac node~\cite{zhu2015ultrafast}. As a result, the hot electrons accumulate in the upper Dirac cone leading to a population inversion. This apparent bottleneck can be attributed to the vanishing phase space at the node. The lifetime of the population inversion is of the order of few ps and the corresponding chemical potential of the photoexcited electrons that can be extracted from the ARPES images is of the order of $100$~meV~\cite{zhu2015ultrafast}. This is the value used for numerical estimates for 3DTIs in Table 1 in the main text. In Ref.~\onlinecite{neupane2015gigantic}, photoexcited states with lifetimes of the order of $\mu$s (accompanied by a $\sim 100$meV shift in the chemical potential) have been observed. We take this result as an indication that long-lived photoexcited states can be realized in 3DTIs. 

\section{Analysis of the contact interaction model}\label{contact_analysis}
For $|\textbf{q}|<<\kappa$ we can see from Eq.~(\ref{eq:coulomb_screened_final}) that the potential is approximately a constant in momentum space
\begin{equation}
V_{\textbf{q}}\approx V_0\equiv \frac{2\pi e^2}{\varepsilon \kappa}.
\label{eq:v0}
\end{equation}
Inserting this expression for the interaction potential to the gap equation, Eq.~(\ref{eq2:gap}), we can see that the gap becomes momentum-independent, 
$\Delta_{\textbf{k}}=\Delta_0$, and satisfies the following equation
\begin{equation}
\Delta_0 = \Delta_0 \frac{\alpha \hbar v}{2\kappa}\int_0^{\Lambda}k dk \frac{\tanh\left( \frac{E_k}{2T}\right)}{E_k}\left[\frac{1-\tanh^2\left(\frac{\mu_{+}}{2T} \right)}{1-\tanh^2\left(\frac{\mu_{+}}{2T} \right)\tanh^2\left( \frac{E_k}{2T}\right)} \right]
\label{eq:gap0}
\end{equation}
where $\alpha=e^2/\varepsilon v$ is the dimensionless coupling constant in the Dirac material, $T$ is the temperature of the electrons and holes, 
and $E_k=\sqrt{(\hbar vk-\mu_{-})^2+\Delta_0^2}$.

The critical temperature for the formation of an excitonic condensate, $T_\mathrm{c}$, can be found by assuming $\Delta_0\neq 0$ and taking $\Delta_0/T\rightarrow 0$. 
We will now analyze the critical temperature in a few different, physically relevant, limits.

First, we consider the limit of perfectly matched chemical potentials, $\mu_{+}=0$. In this limit, assuming $\mu_{-}>2T$, $\mu_{-}/\hbar v \Lambda<<1$, and $T/\hbar v \Lambda<<1$, 
we find the critical temperature can be approximated by:
\begin{equation}
T_\mathrm{c}=C_{-}\sqrt{\mu_{-}(\hbar v \Lambda - \mu_{-})}\exp\left[\frac{\hbar v (\Lambda-2\kappa\alpha^{-1})}{2\mu_{-}}\right]
\label{eq:tc}
\end{equation} 
where $C_{-}=\frac{2}{\pi}e^{\gamma-1}$ where $\gamma\approx0.577216$ is the Euler–Mascheroni constant. From this expression we can identify three distinct cases which depend 
on the competition between the ratio of the screening wavevector to the coupling, $\kappa/\alpha$, and the cutoff wavevector, $\Lambda$: case (i) $\Lambda>2\kappa\alpha^{-1}$; 
case (ii) $\Lambda=2\kappa\alpha^{-1}$; and case (iii) $\Lambda<2\kappa\alpha^{-1}$. 

\begin{figure*}
 \begin{center}
  \centering
  \includegraphics[width=0.97\textwidth]{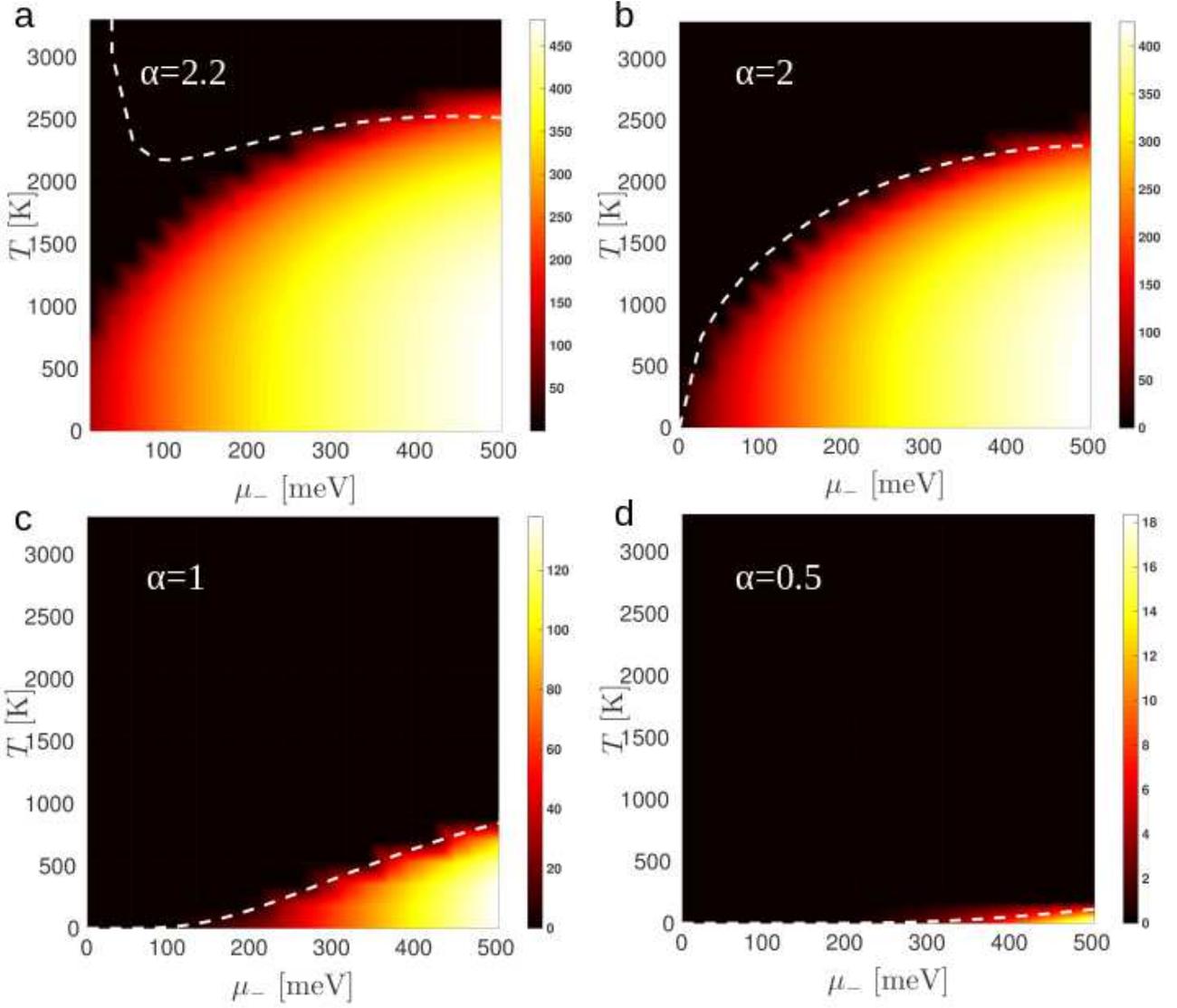}
  \caption{Color plot of the order parameter $\Delta_0$ in the $T$-$\mu_{-}$-plane for $\mu_{+}=0$ computed numerically by solving Eq.~ (\ref{eq:gap0}) 
  self-consistently assuming $\kappa=\Lambda=1/6.58 \text{\AA}^{-1}$ and $v=6.58\text{eV}\cdot\text{\AA}$ for four different values of $\alpha$: (a) $\alpha=2.2$, (b) 
  $\alpha=2$, (c) $\alpha=1$, (d) $\alpha=0.5$. In each case we plot the expression for $T_\mathrm{c}$ using Eq.~(\ref{eq:tc}), shown with dashed lines, and see excellent 
  agreement except for the small $\mu_{-}$ region in (a) due to the exponential enhancement of $T_\mathrm{c}$ as discussed in the text. $\Delta_0$ is in units of meV.}
  \label{fig:T_c_contact}
 \end{center}
\end{figure*}

In case (i) the argument of the exponential is positive and thus $T_\mathrm{c}$ is exponentially enhanced for small $\mu_{-}$. This case is naturally the most favorable for the 
formation of the excitonic condensate which makes sense given that it is the case associated with the limit of strong coupling and weak screening. We should note that in this case, 
for very small values of $\mu_{-}$, Eq.~(\ref{eq:tc}) no longer applies since the exponential enhancement of $T_\mathrm{c}$ will render our assumption $T/\hbar v \Lambda<<1$ invalid. 
This disagreement is demonstrated in Fig.~\ref{fig:T_c_contact}(a). However, as we can see, the expression in Eq.~(\ref{eq:tc}) still agrees qualitatively away from $\mu_{-}=0$. 

In case (ii) the exponential factor is equal to unity and thus $T_\mathrm{c}$ possesses a simple square-root dependence on $\mu_{-}$. In this case any finite value of $\mu_{-}$ will 
lead to a finite $T_\mathrm{c}\sim \sqrt{\mu_{-}}$. In Fig.~\ref{fig:T_c_contact}(b) we demonstrate that this result agrees very well with the numerically computed phase diagram. 

In case (iii) the argument of the exponential is negative and thus $T_\mathrm{c}$ is exponentially suppressed for small $\mu_{-}$. This dependence seems natural since this is the case 
associated with the limit of strong screening and weak coupling. In Figs.~\ref{fig:T_c_contact}(c) and (d) we show phase diagrams for two cases falling into this regime demonstrating 
the exponential suppression of $T_\mathrm{c}$ for small $\mu_{-}$.  

Next, we will discuss what happens to the phase diagram for $\mu_{+}\neq 0$. From Eq.~(\ref{eq:gap0}) we can see that in the limit of $\mu_{+}/2T>>1$ the only solution is $\Delta_0=0$. 
Hence, it is clear that a mismatch between the two chemical potentials ($\mu_e$ and $\mu_h$) is potentially destructive. Allowing for a small but finite mismatch, $\mu_{+}/2T<<1$, we 
can expand Eq.~(\ref{eq:gap0}) in powers of $\mu_{+}/2T$ and we find that the leading order corrections to $T_\mathrm{c}$ are given by:
\begin{equation}\label{Tc_unb_AppB}
T_\mathrm{c}(\mu_{+})\approx T_\mathrm{c}(0)\left(1 - C_{+} \frac{\mu_{+}^2}{4T_\mathrm{c}(0)^2} \right) 
\end{equation}
where $C_{+}=\gamma+\ln{\frac{4}{\pi}}-1/3$ and $T_\mathrm{c}(0)$ is the critical temperature given by Eq.~(\ref{eq:tc}). From this expression we can see that even small 
deviations from $\mu_{+}=0$ will lead to a reduced value of $T_\mathrm{c}$.

\section{Time-evolution of the order parameter}\label{time_evolution}
To account for the transient nature of the excitonic states in a pumped DM, we require a dynamical approach. In this section, we employ semiconductor Bloch equations 
(SBE)~\cite{lindberg1988be,haug2004quantum,stroucken2011optical,malic2011graphene,winzer2013microscopic} to model the dynamics of photoexcited carriers in a pumped DM in the 
presence of interactions. This approach yields a system of differential equations of motion that describe the time evolution of the basic single-particle expectation values, 
namely electron and hole populations, $n_{\textbf{k}}^\mathrm{e/h}=\left\langle \psi_{\mathrm{e/h},\textbf{k}}^{\dagger}\psi_{\mathrm{e/h},\textbf{k}}\right\rangle$, and the 
anomalous correlator (interband polarization) $f_{\textbf{k}}=\left\langle\psi_{\mathrm{e},-\textbf{k}}\psi_{\mathrm{h},\textbf{k}}^{\dagger}\right\rangle$, which is related to 
the order parameter $\Delta_{\mathbf{k}}$ as $\Delta_{\textbf{k}}=\sum_{\textbf{k}^{\prime}}V_{\textbf{k}-\textbf{k}^{\prime}}f_{\textbf{k}^{\prime}}$. The numerical solution of 
the SBE yields the time-evolution of the order parameter and electron and hole occupations. 

To derive the SBE for a pumped DM, we write down equations of motion for $n_{\textbf{k}}^\mathrm{e/h}$ and $f_\mathrm{k}$. For any operator $O$, we have
\begin{eqnarray}
\frac{d\left\langle O\right\rangle}{dt}=\frac{i}{\hbar}\left\langle[H,O]\right\rangle,
\end{eqnarray}
where $H$ is given in Eq.~(\ref{eq:H}) with the interaction term in the mean-field approximation. After computing the commutators of $O$ with each term in the Hamiltonian we 
obtain the following system of equations
%
%\begin{eqnarray}
\begin{equation}
\begin{aligned}
\frac{d n_{\mathbf{k}}^{e}}{dt}&=i\Delta_\mathbf{k}^{*}f_\mathbf{k}^{*}-i\Delta_\mathbf{k} f_\mathbf{k}+\frac{d n_{\mathbf{k}}^{e}}{dt}|_\mathrm{scat},\\
\frac{d n_{-\mathbf{k}}^{h}}{dt}&=i\Delta_\mathbf{k}^{*}f_\mathbf{k}^{*}-i\Delta_\mathbf{k} f_\mathbf{k}+\frac{d n_{\mathbf{k}}^{h}}{dt}|_\mathrm{scat},\\
\frac{d f_{\mathbf{k}}}{dt}&=i(\varepsilon_{\mathbf{k}}^e+\varepsilon_{\mathbf{k}}^h)f_{\mathbf{k}}+i\Delta_\mathbf{k}^{*}(1-n_\mathbf{k}^{e}-n_{-\mathbf{k}}^{h})+\frac{d f_{\mathbf{k}}}{dt}|_\mathrm{scat}.
\label{SBE}
\end{aligned}
\end{equation}
%\end{eqnarray}
%

Dissipation has been incorporated into the equations of motion via phenomenological scattering terms denoted as $d/dt|_\mathrm{scat}$ in Eq.~(\ref{SBE}). 
We take into account two main mechanisms of relaxation, (i) the interband relaxation due to recombination of carriers, which results in a decrease of the electron 
and hole populations and thus the magnitudes of the chemical potentials, and (ii) intraband relaxation due to intraband scattering, which results in thermal equilibration 
of carriers at an instantaneous chemical potential $\mu_{\mathrm{e(h)}}(t)$ and time $t$. These processes are described by relaxation times $T_1$ and $T_1^\prime$, respectively.  
The scattering terms in Eq.~(\ref{SBE}) take on the following form
%
%\begin{eqnarray}
\begin{equation}
\begin{aligned}
\frac{d n_{\mathbf{k}}^{e}}{dt}|_{scat}&=-\frac{n_{\mathbf{k}}^{e}(t)-n_{\mathrm{F}}(\mu^{e}(t))}{T_1^{\prime}}-\frac{n_{\mathbf{k}}^{e}(t)}{T_1},\\
\frac{d f_{\mathbf{k}}}{dt}|_{scat}&=-\frac{f_{\mathbf{k}}(t)}{T_2}.
\end{aligned}
\end{equation}
%\end{eqnarray}

The form of the scattering terms is obtained by coupling the electron and hole subsystems to a pair of featureless (fermionic or bosonic) reservoirs and by subsequently 
integrating out the reservoir degrees of freedom~\cite{stefanucci2013nonequilibrium, goldstein2015photoinduced}. The relaxation terms can be derived microscopically from 
many-particle interactions e.g. in the second-order Born-Markov approximation~\cite{malic2011graphene}. Relaxation dynamics is mainly governed by carrier-carrier and 
carrier-phonon scattering which can contribute to both intraband and interband relaxation.	

\begin{figure*}[ht!]
\begin{center}
\centering
\includegraphics[width=0.99\linewidth,clip=true]{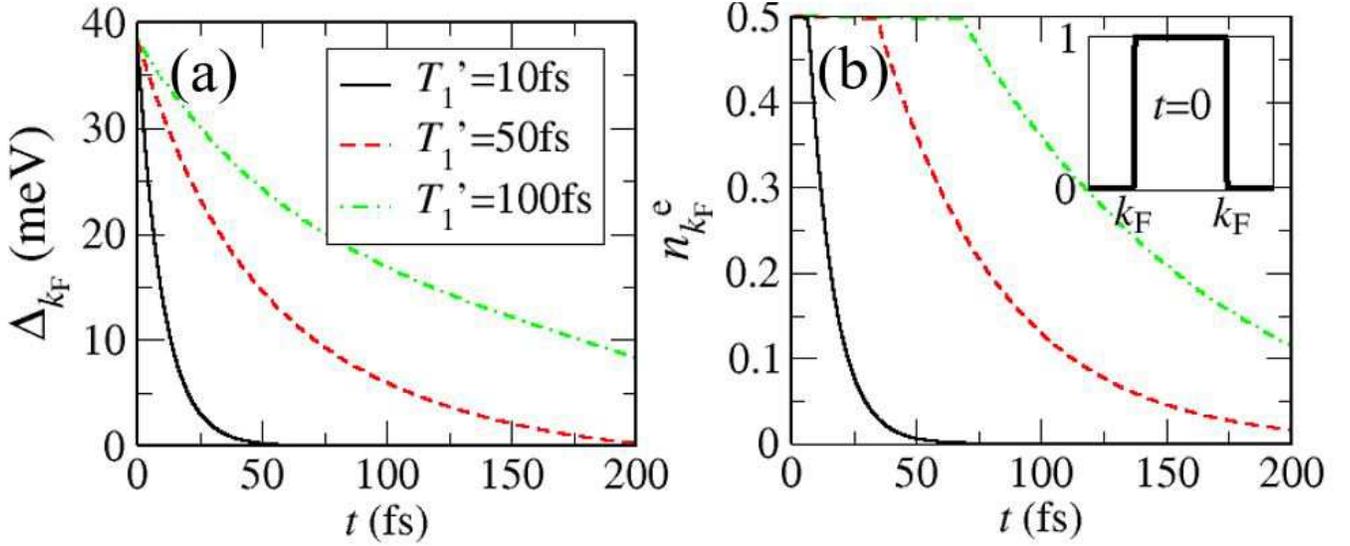}
\caption{Dynamics of (a) the order parameter and (b) electron occupation obtained from numerical integration of the SBE, Eq.~(\ref{SBE}), for $T_1>>T_1^\prime$ with different 
$T_1^\prime$. We use parameters for graphene with $\alpha=0.2$, $T=0$ and $\mu_\mathrm{e}=\mu_\mathrm{h}=500$~meV. The inset in panel (b) shows distribution of $n_{k}^{e}$ at $t=0$.}
\end{center}
\label{fig_dyn}
\end{figure*}

The decay of the excitonic state is governed by the dephasing time $T_2$ related to the scattering times as $T_2^{-1}={T_1^\prime}^{-1}+T_1^{-1}$~\cite{malic2011graphene}. 
Figure~\ref{fig_dyn} shows the time-evolution of the order parameter and electron occupation (the dynamics of electrons and holes is identical). We take as initial state 
the values of the gap and occupations in the quasiequilibrium state at fixed $\mu_\mathrm{e}$ and $\mu_\mathrm{h}$, which are then evolved according to equations of motion. 
In doing so we neglect the ultrafast processes associated with excitation and with building up of the transient inverted population and focus only on the relaxation of the 
transient state towards equilibrium. The gap and the instantaneous chemical potential are calculated self-consistently at each time step. In these simulations we consider a 
regime in which $T_1>>T_1^\prime$ for different $T_1^\prime$ and we limit the total simulation time to a few hundred fs. 

Since all relaxation channels contribute to the dephasing of the the interband polarization $f_{\textbf{k}}$ and hence $\Delta_{\mathbf{k}}$, the lifetime of the 
gapped excitonic state is determined by the shortest of the relaxation times (the largest scattering rate). Experimental results~\cite{gierz2015graphene} and microscopic 
modeling~\cite{winzer2013microscopic} of ultrafast relaxation dynamics in graphene suggest that the Coulomb-induced interband interaction, in particular Auger scattering 
(recombination) has the largest scattering rate and is predominantly responsible for the relaxation of the transient population inversion towards equilibrium within 100-200~fs. 
This gives an estimate for the lifetime of the transient excitonic state. According to recent experiments, lifetimes of the population inversion in 3DTIs are orders of magnitude 
larger ($4$ps-$\mu$s); however, detailed microscopic calculations similar to the ones done for graphene are needed to clarify the relative contributions of different scattering channels.       

\bibliography{Driven_Dirac}

\end{document}